\PassOptionsToPackage{unicode}{hyperref}
\PassOptionsToPackage{hyphens}{url}
\PassOptionsToPackage{dvipsnames,svgnames,x11names}{xcolor}
\documentclass[
  sn-nature,
  11pt,
]{sn-jnl}

\usepackage{amsmath,amssymb}
\usepackage{iftex}
\ifPDFTeX
  \usepackage[T1]{fontenc}
  \usepackage[utf8]{inputenc}
  \usepackage{textcomp} 
\else 
  \usepackage{unicode-math}
  \defaultfontfeatures{Scale=MatchLowercase}
  \defaultfontfeatures[\rmfamily]{Ligatures=TeX,Scale=1}
\fi
\usepackage{lmodern}
\ifPDFTeX\else  
\fi
\IfFileExists{upquote.sty}{\usepackage{upquote}}{}
\IfFileExists{microtype.sty}{
  \usepackage[]{microtype}
  \UseMicrotypeSet[protrusion]{basicmath} 
}{}
\makeatletter
\@ifundefined{KOMAClassName}{
  \IfFileExists{parskip.sty}{%
    \usepackage{parskip}
  }{
    \setlength{\parindent}{0pt}
    \setlength{\parskip}{6pt plus 2pt minus 1pt}}
}{
  \KOMAoptions{parskip=half}}
\makeatother
\usepackage{xcolor}
\makeatletter
\ifx\paragraph\undefined\else
  \let\oldparagraph\paragraph
  \renewcommand{\paragraph}{
    \@ifstar
      \xxxParagraphStar
      \xxxParagraphNoStar
  }
  \newcommand{\xxxParagraphStar}[1]{\oldparagraph*{#1}\mbox{}}
  \newcommand{\xxxParagraphNoStar}[1]{\oldparagraph{#1}\mbox{}}
\fi
\ifx\subparagraph\undefined\else
  \let\oldsubparagraph\subparagraph
  \renewcommand{\subparagraph}{
    \@ifstar
      \xxxSubParagraphStar
      \xxxSubParagraphNoStar
  }
  \newcommand{\xxxSubParagraphStar}[1]{\oldsubparagraph*{#1}\mbox{}}
  \newcommand{\xxxSubParagraphNoStar}[1]{\oldsubparagraph{#1}\mbox{}}
\fi
\makeatother

\usepackage{longtable,booktabs,array}
\usepackage{calc} 
\usepackage{etoolbox}
\makeatletter
\patchcmd\longtable{\par}{\if@noskipsec\mbox{}\fi\par}{}{}
\makeatother
\IfFileExists{footnotehyper.sty}{\usepackage{footnotehyper}}{\usepackage{footnote}}
\makesavenoteenv{longtable}
\usepackage{graphicx}
\makeatletter
\newsavebox\pandoc@box
\newcommand*\pandocbounded[1]{
  \sbox\pandoc@box{#1}%
  \Gscale@div\@tempa{\textheight}{\dimexpr\ht\pandoc@box+\dp\pandoc@box\relax}%
  \Gscale@div\@tempb{\linewidth}{\wd\pandoc@box}%
  \ifdim\@tempb\p@<\@tempa\p@\let\@tempa\@tempb\fi
  \ifdim\@tempa\p@<\p@\scalebox{\@tempa}{\usebox\pandoc@box}%
  \else\usebox{\pandoc@box}%
  \fi%
}
\def\fps@figure{htbp}
\makeatother

\usepackage{graphicx}%
\usepackage{multirow}%
\usepackage{amsmath,amssymb,amsfonts}%
\usepackage{amsthm}%
\usepackage{mathrsfs}%
\usepackage[title]{appendix}%
\usepackage{xcolor}%
\usepackage{textcomp}%
\usepackage{manyfoot}%
\usepackage{booktabs}%
\usepackage{algorithm}%
\usepackage{algorithmicx}%
\usepackage{algpseudocode}%
\usepackage{listings}%

\usepackage{lipsum} \usepackage{libertine}
\makeatletter
\@ifpackageloaded{caption}{}{\usepackage{caption}}
\AtBeginDocument{%
\ifdefined\contentsname
  \renewcommand*\contentsname{Table of contents}
\else
  \newcommand\contentsname{Table of contents}
\fi
\ifdefined\listfigurename
  \renewcommand*\listfigurename{List of Figures}
\else
  \newcommand\listfigurename{List of Figures}
\fi
\ifdefined\listtablename
  \renewcommand*\listtablename{List of Tables}
\else
  \newcommand\listtablename{List of Tables}
\fi
\ifdefined\figurename
  \renewcommand*\figurename{Figure}
\else
  \newcommand\figurename{Figure}
\fi
\ifdefined\tablename
  \renewcommand*\tablename{Table}
\else
  \newcommand\tablename{Table}
\fi
}
\@ifpackageloaded{float}{}{\usepackage{float}}
\floatstyle{ruled}
\@ifundefined{c@chapter}{\newfloat{codelisting}{h}{lop}}{\newfloat{codelisting}{h}{lop}[chapter]}
\floatname{codelisting}{Listing}

\makeatother
\makeatletter
\@ifpackageloaded{caption}{}{\usepackage{caption}}
\@ifpackageloaded{subcaption}{}{\usepackage{subcaption}}
\makeatother

\usepackage{bookmark}

\IfFileExists{xurl.sty}{\usepackage{xurl}}{} 
\hypersetup{
  pdftitle={Producing population-level estimates of internal displacement in Ukraine using GPS mobile phone data},
  pdfauthor={Rodgers Iradukunda; Francisco Rowe; Elisabetta Pietrostefani},
  pdfkeywords={internal population displacement, GSP mobile phone
data, Ukraine},
  colorlinks=true,
  linkcolor={blue},
  filecolor={Maroon},
  citecolor={Blue},
  urlcolor={Blue},
  pdfcreator={LaTeX via pandoc}}

\title[\textbf{Producing population-level estimates of internal
displacement in Ukraine using GPS mobile phone data}]{\textbf{Producing
population-level estimates of internal displacement in Ukraine using GPS
mobile phone data}}

\author[1]{\fnm{Rodgers} \sur{Iradukunda}}\author*[1]{\fnm{Francisco} \sur{Rowe}}\email{f.rowe-gonzalez@liverpool.ac.uk}\author[1]{\fnm{Elisabetta} \sur{Pietrostefani}}
\affil[1]{\orgdiv{Geographic Data Science Lab, Department of Geography
and Planning}, \orgname{University of Liverpool, Liverpool}}

\abstract{Nearly 110 million people are forcibly displaced people
worldwide. However, estimating the scale and patterns of internally
displaced persons in real time, and developing appropriate policy
responses, remain hindered by traditional data streams. They are
infrequently updated, costly and slow. Mobile phone location data can
overcome these limitations, but only represent a population segment.
Drawing on an anonymised large-scale, high-frequency dataset of
locations from 25 million mobile devices, we propose an approach to
leverage mobile phone data and produce population-level estimates of
internal displacement. We use this approach to quantify the extent, pace
and geographic patterns of internal displacement in Ukraine during the
early stages of the Russian invasion in 2022. Our results produce
reliable population-level estimates, enabling real-time monitoring of
internal displacement at detailed spatio-temporal resolutions. Accurate
estimations are crucial to support timely and effective humanitarian and
disaster management responses, prioritising resources where they are
most needed.}


\begin{document}
\maketitle

\newpage

The forced displacement of individuals, including refugees,
asylum-seekers and internally displaced people (IDP), creates
considerable humanitarian, social and economic costs
\citep{blattman2010, unhcr2023}. Recent estimates indicates that the
number of forcibly displaced populations has significantly grown as
result of persecution, conflict, violence, human rights violations and
disasters \citep{idmc2024}. As of June 2023, the United Nations High
Commissioner for Refugees (UNHCR) estimated 110 million of forcibly
displaced people worldwide, with the number of IDP (62.5 million)
accounting for the largest share of these displacements
\citep{UNHCR_mid2023}. The Russian full-scale invasion of Ukraine is
estimated to have created the fastest global displacement crisis, and
one of the largest, since the Second World War \citep{unhcr2023}.

Forcibly displaced population data are key to inform operational plans,
humanitarian responses and long-term policy making. By understanding the
scale and locations where people are forcibly fleeing and the extent of
their return, government agencies, aid organisations and local community
groups can better prioritise and allocate resources and services where
they are most needed in the required quantities \citep{idmc2024}. Highly
granular geographical data tracking population displacements in real
time are therefore critical to support these efforts
\citep{rowe2022, gonzález-leonardo2024}.

Traditional data systems are constrained to render information at such
high temporal and geographical resolution and speed. Over the years,
UNHCR and the Internal Displacement Monitoring Centre (IDMC) have made
significant efforts triangulating various data sources to improve and
deliver global databases that enable the monitoring and management of
forced population displacements \citep{idmc2024}. However, they have
also identified persistent challenges in the production of reliable
estimates of forcibly displaced populations
\citep{sarzin2017stocktaking, tai2022}. Traditional data systems are not
regularly updated, costly and characterised by slow data collection and
release \citep{rowe2023}. Particularly in conflict areas, humanitarian
partners and data collectors often face access restrictions due to
violence and insecurity preventing data gathering \citep{UNHCR_mid2023}.
Data streams may also have gaps collecting data on displacement during
short-term evacuations or spontaneous movements resulting from conflict
and violence \citep{guideto2019, drouhot2022}. The danger and
challenging nature of field work in conflict zones can also disrupt
continuous engagement in data collection by humanitarian and development
agencies \citep{understa2011, salehyan2015}.

Novel digital footprint data have emerged as a key source of information
offering an opportunity to capture human population movements at highly
granular geographical and temporal scales
\citep{checchi2013validity, rowe2023}. These data are automatically and
continuously generated avoiding exposure of data collectors to hazardous
areas and minimising potential data gaps \citep{guideto2019}. Mobile
phone location data have increasingly been used to monitor population
movements during crises, particularly measuring exposure to ambient
pollutant exposure \citep{nyhan2016exposure, dewulf2016dynamic},
transport patterns \citep{huang2019transport}, recreational behaviour
\citep{kim2023mobile}, disaster-induced displacement (e.g.~flooding and
earthquakes) \citep{lu2016unveiling} and the spread of diseases -
notably during the COVID-19 pandemic \citep{grantz2020use}. Yet, limited
work has been undertaken to estimate the scale and patterns of IDP using
mobile phone data.

Additionally, differences in the access and use of mobile phone
technology and applications used to collect location data prevent the
production of reliable population-level mobility estimates. Most
existing work based on mobile phone data has thus constrained to offer
rough signals about population movements (e.g.~spatial concentration),
trends (e.g.~increasing) and changes (e.g.~low to high)
\citep{rowe2023}. For example, a recent study \citep{shibuya2024} used
GPS data to offer a granular representation of the geographic and
temporal patterns and trends in population displacement in Ukraine, but
the reported estimates correspond changes in the number of unique mobile
phone devices. They do not represent population-level figures. Estimates
need to be adjusted for biases to make them representative of the full
population.

To address these gaps, we propose an approach to produce high-frequency
population-level estimates of internal displacement drawing on location
data from 25 million unique devices. Our approach adapts Leasure's et
al.~approach \citep{leasure2023nowcasting} of bias data correction to
GPS human mobility data. Leasure's et al.~developed a method of bias
data correction for Facebook users data extracted from Facebook
Marketing Advertising platform to produce estimates of population
displacement in Ukraine. A key limitation is that the Facebook Marketing
Advertising data represent counts of active Facebook users in an area at
coarse geographical scales. Thus, while they can be used to generate
estimates of local population counts, they do not allow the estimation
of displacement flows to identify displacement routes, their origins and
destinations, limiting their insights to support humanitarian
operations. Rather than as a simple application, we see the adaptation
of Leasure's et al.~approach to GPS data as a major extension of the
approach. It requires the engineering of raw GPS data (only containing a
device identifier, a timestamp, longitude and latitude) to identify home
locations before and after the start of the war, and then determine
changes in usual place of residence.

We make two substantive contributions. Our first contribution is
methodological and illustrates how high-frequency footprint data can
enable the generation of population-level estimates of internal
displacement correcting for differences in mobile phone-derived and
actual population counts, moving beyond providing rough signals. Most
prior work leveraging on digital footprint data to estimate population
displacement relies on social media or call detail records, with
location being inferred resulting in reduced precision
\citep{ranjan2012, zhao2016, pestre2019}. We use data collected via GPS
technology which provides greater precision data on location
\citep{grantz2020}.

Our second contribution is to provide evidence of the scale and spatial
patterns of population displacement in Ukraine during the first year of
the invasion. The Russian full-scale invasion of Ukraine has created the
fastest global displacement crisis, and one of the largest, since the
Second World War \citep{unhcr2023}. Recent estimates suggest that nearly
one-third of Ukrainian residents are estimated to have been forced from
their homes \citep{unhcr2023}. As of 25 September 2023, 3.67 million
people were estimated to have been displaced internally within Ukrainian
borders \citep{IOM2022}. These estimates are based on a random digit
dial telephone survey aiming at generating a nationally representative
sample of 2,000 individuals at each monthly round \citep{IOM2022}. While
consistent with high frequency estimates based on Facebook data
\citep{leasure2023nowcasting}, these estimates cannot deliver
population-level estimates of population displacement for subnational
areas at high granularity, or high-temporal frequency. Our approach
offers high frequency population displacement estimates to complement
data derived from traditional data streams.

\section{Results}\label{results}

\subsection{Estimating the extent of internal population
displacement}\label{estimating-the-extent-of-internal-population-displacement}

\begin{figure}

\begin{minipage}{\linewidth}

\pandocbounded{\includegraphics[keepaspectratio]{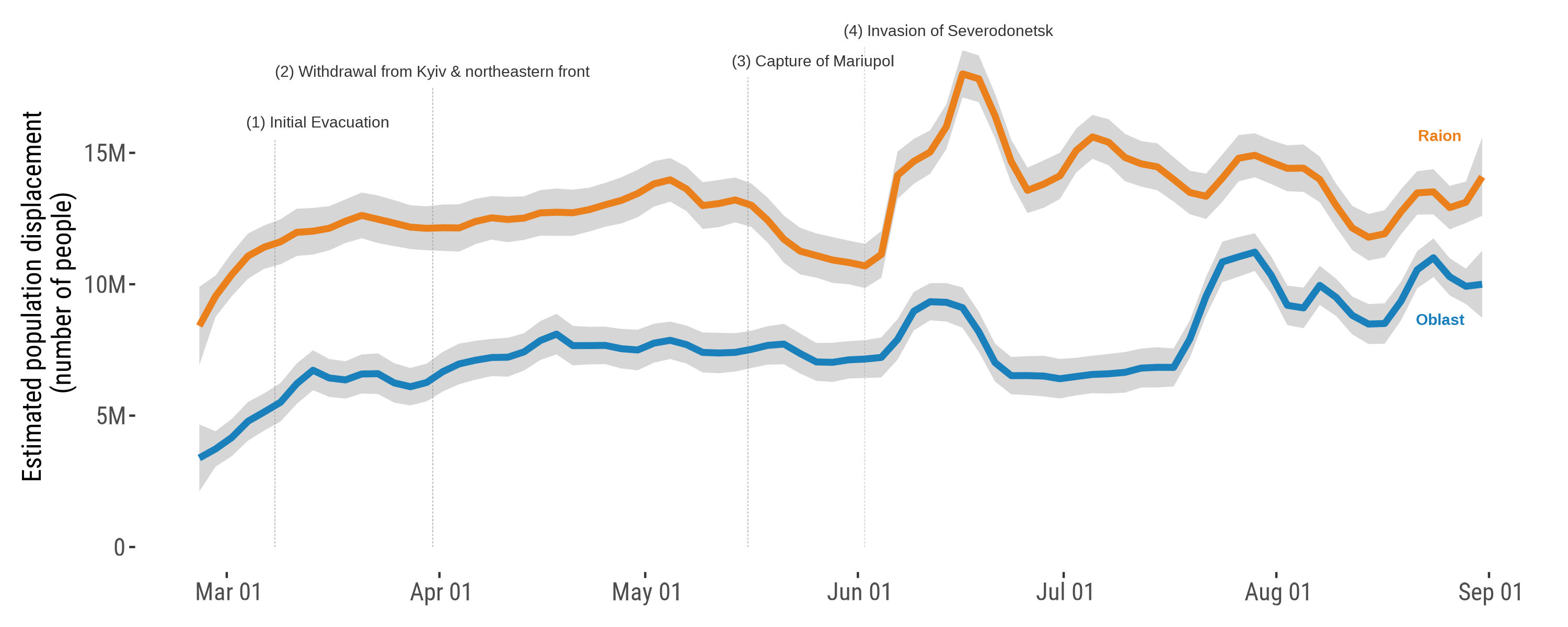}}

\end{minipage}%

\caption{\label{fig-popdisplacement}\textbf{Estimated daily number of
displaced population. February to August 2022.} The number of displaced
people is estimated as the difference between the population in a region
for a given day after the start of the war and the population before the
war in 2020 - see Section~\ref{sec-methods2.2}.}

\end{figure}

We first estimate the extent of daily internal population displacement
at the oblast and raion level (Figure~\ref{fig-popdisplacement}). We
estimate that over 5 million people were internally displaced from their
oblast of residence by April 2022 reaching an average of about 10
million in late July and August 2022. Figure~\ref{fig-popdisplacement}
reveals a drop in population displacement during mid-June and mid-July,
coinciding with a pattern of return displacements primarily to the
cities of Kiev and Kharkhiv (see Section~\ref{sec-return}). In addition
to return movements, subsequently higher but fluctuating levels of
movement after mid July seem to reflect the shifting dynamics of the
armed conflict towards southeastern Ukraine where war fire intensified
during this period \citep{walker2024}.

Our contribution is to generate geographically granular estimates of
internal displacement at the raion level leveraging the high spatial
precision of GPS data. As anticipated, the levels of raion-level
displacement consistently exceeds those of oblast-level displament as
they reflect movements that cannot be captured at higher levels of
spatial aggregation: raions within the same oblast's boundaries
capturing the fact that most displacement tends to occur over short
distances. Our raion-level estimates indicate a rise and peak of over 17
million displaced people in mid June 2022 following the start of the
Russian invasion of Severodonetsk. Around 90\% of the buildings and
infrastructure is estimated to have been destroyed or damaged after the
capture of Severodonetsk \citep{yana2021}. From mid July, our estimates
indicate a rise in population displacement at the oblast level, but such
increase is not reflected at the raion level, indicating that the most
displacement that took place during this time tended to occur over long
distances involving a cross of oblast boundaries (see Fig.1 in the
Supplementary Material (SM) displaying distance distributions).

Our findings are consistent with existing estimates. We compare our
oblast-level displacement estimates with existing estimates derived from
an United Nations - International Organization for Migration (IOM)
survey \citep{IOM2022} and Facebook data \citep{leasure2023nowcasting}
(see Section~\ref{sec-methods2.2}, Table 1 and Figures 2-4 in SM). The
shape of the temporal evolution of population displacement is remarkably
consistent. Though, we identify some discrepancies. Our estimates tend
to be higher than those produced by Leasure et al.~by approximately 250
thousand people across the time series. The difference can be explained
by Leasure et al.'s estimates are affected by power outages in the
Donetsk and Luhansk regions resulting in zero or small numbers for
various dates \citep{rowe2022b, leasure2023nowcasting} (see Figure 3 in
SM). Similarly, our oblast-level estimates are noticable greater than
the IOM figures in June and August. We assume that this is because our
estimates include data from Crimea, and there was significant movement
from and to Crimea to Russian-occupied Ukrainian territory and Russia
during these months \citep{walker2024}. This is as Russia started a
``volunteer mobilisation'' and deployed new troops and logistics to
support an a frontline extending from Zaporizhzhia to Kherson, along the
Dnieper River \citep{walker2024}. If we exclude Crimea, our estimates
are much closer to IOM and Leasure et al.'s estimates (see Table 1 in
SM).

\subsection{Identifying the main origins and
destinations}\label{sec-odm}

\begin{figure}

\begin{minipage}{\linewidth}

\includegraphics[width=\linewidth,height=0.5\textheight,keepaspectratio]{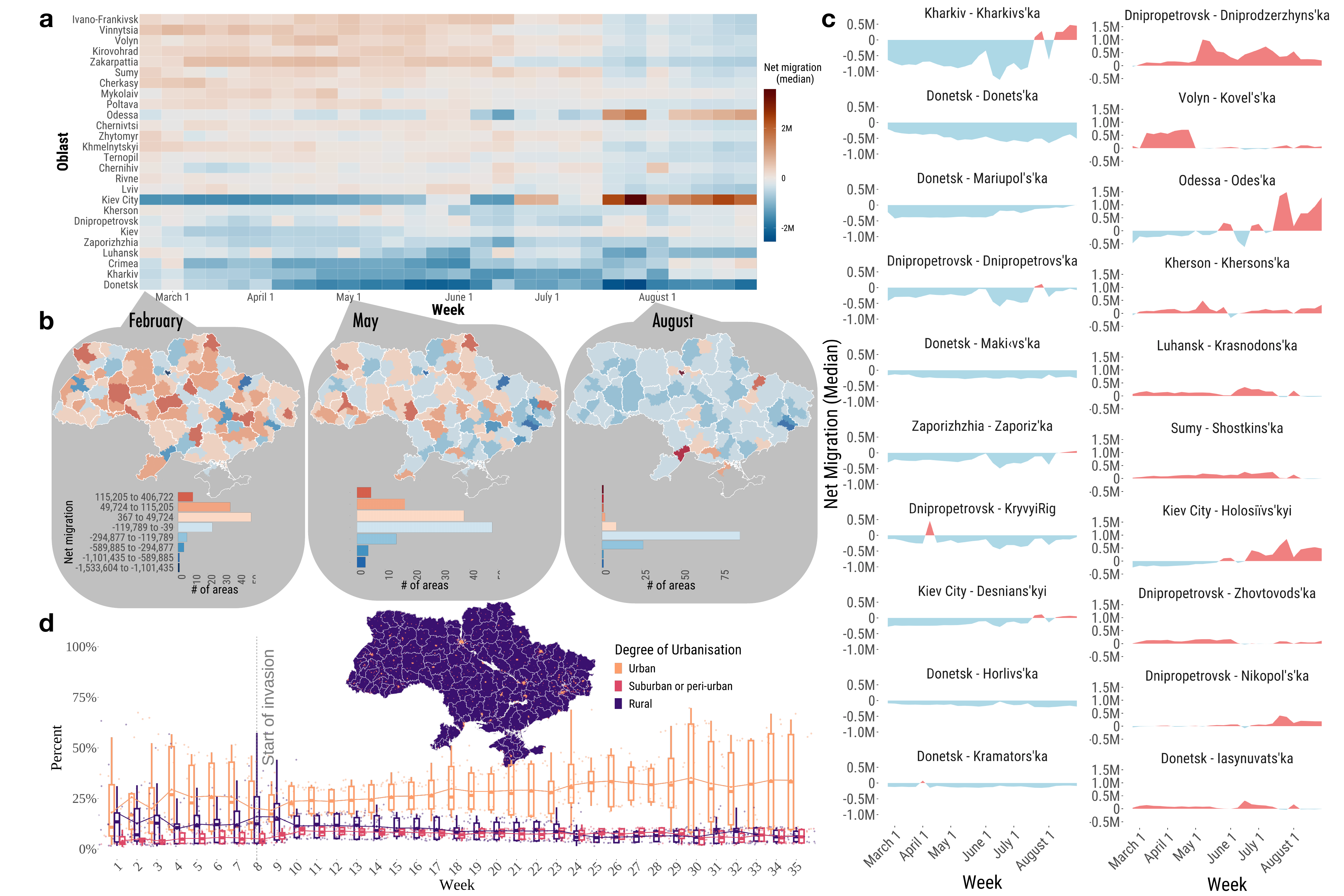}

\end{minipage}%

\caption{\label{fig-heatmaps}\textbf{Net migration count by oblast and
raions, February to August 2022.} \textbf{a.} Weekly median net
migration by oblast. \textbf{b.} Monthly median net migration across
raions. \textbf{c.} Top ten raions with the largest cumulative net
negative (in blue) and positive (in red) migration balance organised
from the largest to the smallest. \textbf{d.} Changes in the share of
estimated population across the urban-rural hierarchy.}

\end{figure}

We then examine the net balance of internal population displacements
resulting inflows minus outflows, to identify the main areas losing and
gaining population through these displacements. As expected,
Figure~\ref{fig-heatmaps}a reveals that Kiev City was the main area
losing population at the start of the war between late February and
early May before recording large positive net balances of over 2 million
people. These gains seem to echo large-scale return population movements
as Russian troops withdrew from the outskirts of Kiev City and focused
on the eastern and southern regions of Ukraine, particularly Donetsk,
Kharkiv, Crimea and Luhansk (Figure~\ref{fig-heatmaps}b). Reflecting the
geographic concentration of military ground forces, these frontline
eastern and southern regions registered a consistent pattern of
population losses between March and August. Population losses are
particularly prominent in Donetsk where the estimated losses exceeded 2
million people in late July and early August 2022. To a lesser extent,
Odessa also displays a negative albeit moderate balance of population
displacements during the early months of the invasion as Russia had a
naval blockade on Ukrainian ports.

At the same time, Figure~\ref{fig-heatmaps}a reveals that western,
central and central-south areas tended to gain population during the
early months of the invasion between February and June 2022. These areas
include oblasts close to the border with Poland, Slovakia, Hungary,
Romania and Moldova, such as Ivano-Frankivs'k, Vinnytsya, Volyn and
Zakarpattia probably serving as transit centres for international
crossings and humanitarian assistance. Kirovohrad also shows
considerable positive population balances over the early months of the
war, most likely receiving population from frontline areas in eastern
parts of Ukraine. Figure~\ref{fig-heatmaps}a shows that most of these
areas have tended to experience population losses as Kiev City and
Odessa record positive population balances from late July.

These aggregate patterns of population displacement conceal the local
concentration of net population losses and gains across raions.
Figure~\ref{fig-heatmaps}c reports the net balance of population
displacements over time for the ten raions with the largest cumulative
losses and gains between February and August 2022. It reveals that
Kharkiv remained the raion with the largest cumulative loss of
population since the start of the war at least until August 2022, but it
reported positive balances as Ukrainian forces launched a
counteroffensive and liberated major settlements in the Kharkiv oblast
in late July and August 2022. The oblast of Donest'k seems to congregate
the raions with the greatest population losses, reflecting the
concentration of frontline activity in raions, such as Donets'ka,
Mariupol's'ka and Makivs'ka.

On the other hand, Figure~\ref{fig-heatmaps}c reveals that raions within
the oblasts of Dnipropetrovsk, Kiev City and Donetsk recorded the
largest cumulative net migration gains at times when these oblasts
recorded moderate overall net migration losses
(Figure~\ref{fig-heatmaps}a). The raions of Dniprodzerzhyns'ka,
Zhotovods'ka and Nikopol's'ka all registered large cumulative population
gains through net migration from February to August 2022 despite
systematic moderate overall negative migration balances in the oblast of
Dnipropetrovs'k. Similarly, the raion of lasynuvats'ka in Donestk
recorded a large cumulative net migration gain despite this being the
oblast with the largest negative migration balances. These results
suggest that people tended to move locally to neighbouring areas, or
were unable to afford moving to more distant locations in western
Ukraine (see Figure 1 in SM).

Additionally, mapping the patterns of net migration
(Figure~\ref{fig-heatmaps}b and Figure~\ref{fig-heatmaps}d) reveals the
increasing prevalence of population loss through net migration in
Ukraine, particularly in less populated areas. In early weeks of the
invasion in February, negative net migration balances concentrated in
urban centres, especially Kiev and Khakiv. As the conflict evolved, net
migration losses seem to have expanded to most of the country
prominently reducing the relative national share of population in very
low density and low density rural areas (Figure~\ref{fig-heatmaps}d).
These reductions in sparsely populous areas appear to have been mirrored
by a growing national share of population in urban centres, with Kiev
and Odessa acting as the major centres of population attraction in
August (Figure~\ref{fig-heatmaps}d).

\subsection{Return movements}\label{sec-return}

Understanding the scale and pace of return movement to residential areas
in conflict zones after a period of displacement is also important to
shape and support humanitarian assistance, successful reintegration,
mental health and community rebuilding programmes
\citep{iom2023-return}. Understanding return movements enables more
efficient resource allocation prioritising areas for infrastructure
reconstruction and service delivery \citep{iom2023-return}. IOM
estimated that 6 million people had returned to their usual place of
residence in Ukraine by August 23 2022 following a two-week period
elsewhere in the country \citep{UNHCR_mid2023}. At the time of writing,
the most recent IOM estimate puts this figure at 4.7 million returnees
in April 11 2024, 14.2 per cent of whom returned from abroad
\citep{iom2024-return}. These estimates are derived from a survey of 20
thousand people, with follow-ups to 1,638 individuals identified as
returnees \citep{iom2024-return}. The proportion of returnees for each
oblast is computed and multiplied by the total population in Ukraine to
derive return estimates. Returnees are identified as those respondents
who spent a two-week period away from their place of residence.

\begin{figure}

\begin{minipage}{\linewidth}

\includegraphics[width=\linewidth,height=0.5\textheight,keepaspectratio]{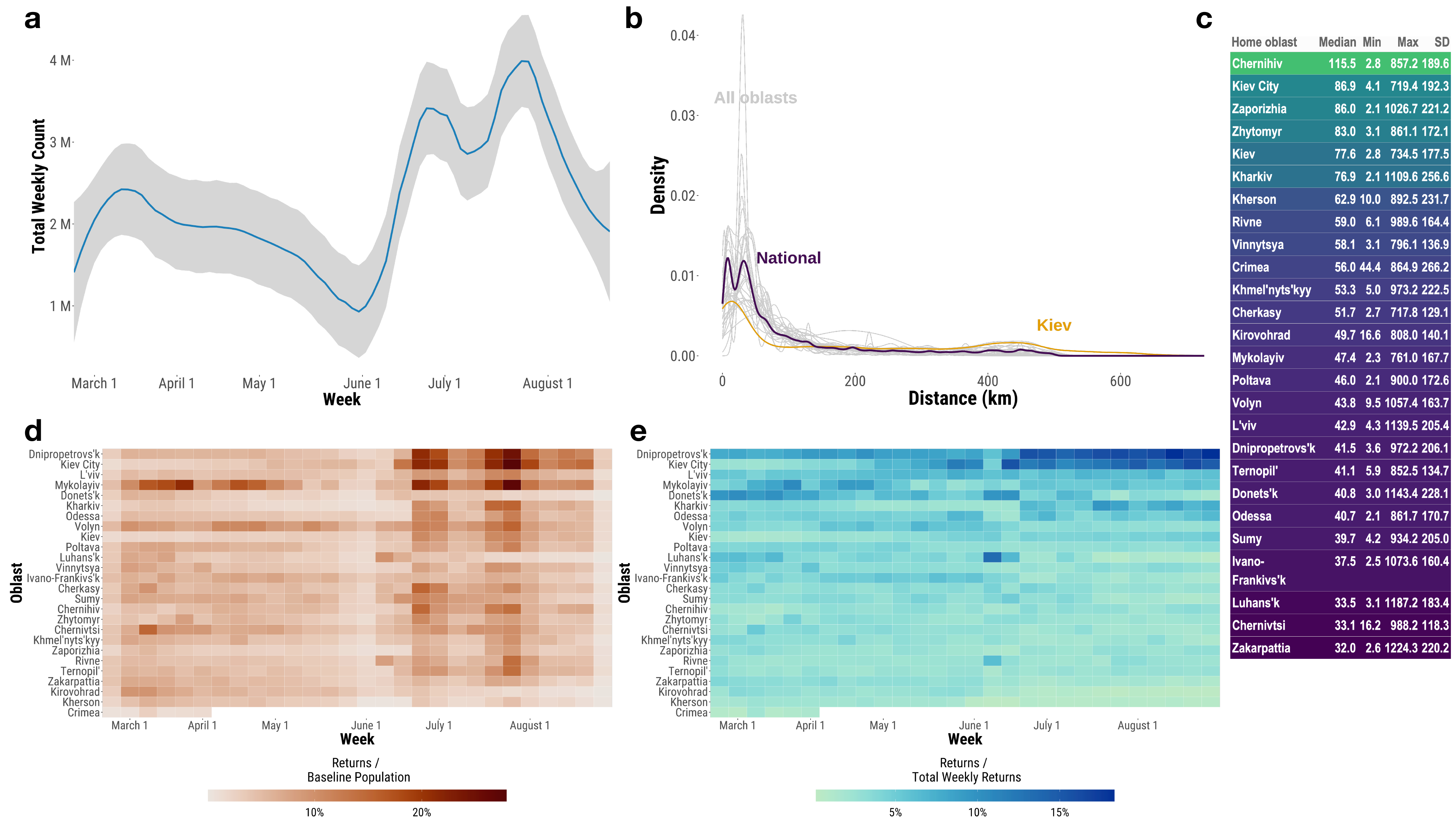}

\end{minipage}%

\caption{\label{fig-returns}\textbf{Return movements, February to August
2022.} \textbf{a.} Weekly total number of returns. Local polynomial
regression modelling was used to build 95\% confidence intervals.
\textbf{b.} Distribution of distance truncated to display return
movements below 700km. \textbf{c.} Median, minimum (min), maximum (max)
and standard deviation (SD) distance in km, oblast. \textbf{d.} Per cent
of returns over the total baseline population for individual oblasts
before the start of the war. \textbf{e.} Per cent of returns over the
total number of returns for individual oblasts.}

\end{figure}%

Using our methodology (see Section~\ref{sec-methods2.2}), we generate
estimates and expand this evidence providing information on the spatial
patterns, distance and pace of return movements
(Figure~\ref{fig-returns}a-e). Our estimates indicate that just over 2
million people had returned to their place of residence before February
24 during the week commencing August 22, 2022
(Figure~\ref{fig-returns}a). Our estimates also showed considerable
fluctuations over time, reflecting that some of displacements and
returns may be temporary. Some people may return to their home location
after spending a short period of time elsewhere. Some may return to
check on close relatives and friends, examine local livability and
recover belongings, and then leave. Our estimates indicate that the
average number of weeks associated with return moves to a home location
is nine weeks (see Figure 6 in SM). The fluctuations observed in our
estimates also reflect the fact that we are unable to follow the same
mobile phone devices for the entire period of analysis. We observe some
individual returns to the same location, but are unable to identify
their location in subsequent periods. Crimea is a good example as we
could only identify returns until the first week of April but not
thereafter (Figure~\ref{fig-returns}d-e).

Figure~\ref{fig-returns}d-e reveal a differentiated rate of return
movements across oblasts. Dnipropetrovs'k and Kiev City display higher
proportions of return movements relative to their populations before the
start of the armed conflict, and to the total weekly number of returns
across Ukraine. L'viv and Mykolaiv record high rates of return movement
likely reflecting their role as transit points, food, temporary shelter
and accommodation centres for refugees, IDP and troops
\citep{gonzález-leonardo2024}. Kherson and Crimea register the lowest
number of returns. As indicated above, no returns were recorded for
Crimea after April 2022, and Kherson remained under Russian occupation
during our period of analysis.

Return movements tend to occur over relatively short distances
(Figure~\ref{fig-returns}b-c). The median distance of return moves
between oblasts is less than 100km suggesting that most IDP tend to stay
relatively close to their home location. Global estimates indicate that
a median distance of less than 100km for internal migration moves is
common \citep{stillwell2016}. However, a wide variation exists as a
function of the place of residence. IDP seem to be willing to travel
longer distances to return Chernivhiv, Kiev City and Zaporizhia than to
Luhans'k, Chernivtsi and Zakarpattia. Chernivhiv and Kiev City recorded
large flows of return movements as Russian troops withdrew from northern
areas of Ukraine and intensifies their war effort on eastern and
southern parts of the country.

\section{Discussion}\label{discussion}

We developed an approach to produce highly granular temporal and spatial
population-level estimates to monitor the extent and geographic patterns
of population displacement in disaster areas drawing on a large dataset
of GPS location data from mobile phone devices. Highly granular data of
internal displacement is essential for real-time monitoring to support
disaster relief and management efforts. Traditional data streams are
limited in their ability to generate such granular information in real
time during times of conflict or natural disasters. Focusing on the
unfolding invasion of Ukraine, we estimated that an increasing number of
people were displaced from their place of residence, with an average of
11 million people being displaced from their Oblast of residence and
over 15 million at the raion level at the start of the Battle of Bakhmut
in early August 2022.

We provided evidence indicating that urban centres were the predominant
locations of population displacement during the early months of the
invasion, with Kiev as the primary origin reporting net migration losses
of approximately 2 million. As the conflict progressed in 2022, we
showed widespread population losses through internal displacements.
Proportionally the share of population in low density rural areas has
reduced mirroring a larger share of population in urban centres. We
showed a systematic increase in the number of return movements to Kiev
City following the withdrawal of Russian forces from the northern and
western front of the city, with frontline areas continuing to lose
population throughout the conflict.

Our work complements existing efforts to generate rapid response
estimates. To estimate population displacement in Ukraine, the IOM
designed a random digit dial telephone survey to produce a nationally
representative sample of 2,000 individuals during each monthly round
\citep{IOM2022r1}. However, this method of data collection is unable to:
(i) generate population level estimates to make inferences of the
geographic patterns of population displacement; (ii) offer temporally
granular frequency estimates (e.g.~daily or weekly) to monitor rapidly
changing population dynamics; or, (iii) produce high spatial resolution
counts to identify areas of humanitarian assistance with high precision
\citep{leasure2023nowcasting}. Prior work has explored the use of
location data from mobile and social media data to address these issues
\citep{graells-garrido2021, gonzález-leonardo2024}. Yet these efforts
have been restricted to provide rough signals of population movements
indicating the direction of trends, spatial patterns and changes of
population flows over time \citep{rowe2023}.

By using pre-conflict population data, we contributed to an approach
that is capable to adjust location data from mobile phone users, moving
away from offering rough signals, to provide estimates of the extent of
population movements. Our approach also has the capacity to provide
real-time monitoring of population displacement at highly temporally and
spatially adaptable resolutions. Our approach can thus complement
existing data resources aiming to provide a national-scale estimate of
population movement. In fact, our estimates aggregated at the national
level were consistent with those derived from the IOM telephone surveys
\citep{IOM2022r1} and social media data \citep{leasure2023nowcasting}.
The triangulation of estimates across these sources helps build
confidence in the official UN estimates, but also on estimates
leveraging innovative data.

We generated population-level estimates using smartphone location data.
However, validating the resulting spatially granular estimates remains a
significant challenge. Normally no comparable estimates exist to
evaluate the extent to which they capture the facts on the ground. That
is the reason why they are produced in the first place. Future efforts
could thus concentrate on making available a repository of high quality
datasets, such as data from comprehensive population registry or
administrative sources that can be used to assess the accuracy of
population-level estimates derived from digital trace data, such as
smartphone data.

We are unable to characterise the population being displaced or their
underpinning reasons using smartphone data. As most digitally generated
data, these data only offer location-time information They do not
provide socio-demographic information about users or their motivations.
As such, we cannot identify the socio-demographic profile of displaced
individuals or why they move; yet, this information is critical to
deliver an appropriate humanitarian response. To tackle this, future
work could assess the integration of area-level data of the resident
population with highly granular displacement estimates derived from GPS
location data to more accurately capture the socio-demographic profile
of displaced communities, and surveys collecting information on why
people move.

We cannot discern between permanent and semi-permanent returns. We can
infer returns if individuals are back to their place of residence
recorded before the start of the war. However, the records of individual
mobile devices offer a rather irregular longitudinal sequence of
locations to confidently determine the time they remained in their place
of residence observed before the start of the war. Future work could
seek to secure a data over a longer time frame which may provide a
larger set of locations over time to distinguish between permanent and
semi-permanent returns.

\section{Methods}\label{methods}

\subsection{Data sources}\label{sec-data-sources}

\textbf{Global Positioning System (GPS) location data.} The primary
source consists of GPS location data from 25 million unique mobile phone
devices. The data include daily GPS locations (longitude and latitude)
in Ukraine, their accuracy and time stamps from January 1st 2022 to
August 31st 2022. Data from digital mobile phone applications are known
to contain biases as they typically represent the behaviour of a segment
of the population \citep{rowe2023}. To mitigate any potential biases
from the use of information from a single source, we use data collected
from a range of mobile applications comprising a variety of users and
purposes. Ethical considerations prevent us from identifying these
applications. The data were obtained from a data vendor.

We process the data to identify unique devices with locations recorded
before (January 1 to February 24, 2022) and after (February 25 to August
31, 2022) the start of the escalation of the Ukraine-Russia conflict. We
identify 17 million devices (approximately 70\% of the total) with
locations before the February 24, and 13 million (approximately 55\% of
the total) with recorded locations after the escalation of the conflict.
We identify 6 million devices with recorded location information for
both before and after the full-scale invasion on February 24, 2022.

To further analyse the data, we apply three main procedures. First, we
apply a point-to-polygon spatial join to assign latitude and longitude
coordinates to administrative boundaries (raions and oblasts) and
settlement area type as defined by the Global Human Settlement Layer
(GHSL) - see description below in Geographic data. Second, we convert
UNIX time stamps into local Ukrainian date and time. Third, we infer
individual home locations for each mobile phone device. For this, we
follow the UN guidelines on official statistics using mobile phone data
\citep{Araietal2022}, and define home location as the place where a
mobile phone is recorded most of the time during night (i.e.~between 7pm
and 5.59am). We calculated the number of days a user's mobile phone
device was detected in the same location during these nighttime hours.
We consider the location where a user's mobile phone device was recorded
for more than 50\% of their time as their place of usual
residence\citep{Araietal2022}.

\textbf{Baseline population data.} We use \(100m^{2}\) gridded
population data to establish the baseline population before the conflict
in Ukraine in 2020 \citep{worldpopukraine}. We utilise unconstrained
population estimates from
\href{https://www.worldpop.org/}{worldpop.org}. These were the most
up-to-date population estimates available for our analysis. We spatially
aggregate the WorldPop population counts to create baseline population
datasets at the raion and oblast levels. These population estimates are
then used to derive population-level estimates of internal displacement,
as described below.

\textbf{Refugee data.} We use United Nations High Commissioner for
Refugees (UNHCR) daily counts of people entering and leaving Ukraine
\citep{OPD}. We accessed these data from an archived version of the
UNHCR website available via \href{https://web.archive.org/}{The Internet
Archive}. These data included daily cross-border movement records from
the start of the full-scale invasion in Ukraine until August 16 2022. We
calculate a cumulative net count by subtracting the number of people
entering Ukraine from those leaving the country. We use this count to
more accurately estimate the number of internally displaced people in
Ukraine by discounting the number of people who moved overseas from the
baseline population.

\textbf{Geographic data.} We use data from two sources. First, we use
geopackages containing the administrative boundaries of Ukraine,
particularly raions and oblasts. We draw on geospatial vector data from
the United Nations Office for the Coordination of Humanitarian Affairs
(OCHA) \href{https://data.humdata.org/dataset?}{Humanitarian Data
Exchange (HDX) data portal} and \href{https://gadm.org/}{Global
Administrative Areas} (GADM) \citep{GADM}. We first spatially join our
GPS mobility data with GADM raion and oblast boundaries. GADM boundaries
contain 629 raions and 26 oblasts. We then aggregate these raions based
on HDX raion boundaries which correspond to the offically recognised
administrative boundaries in Ukraine.

We also use the degree of urbanisation classification from the
\href{https://human-settlement.emergency.copernicus.eu/}{GHSL} to
determine the type of settlement areas of IDP, both origins and
destinations. We reclassify the seven original categories to identify
three types of areas: urban (dense urban cluster and urban centre),
suburban or peri-urban (suburban and semi-dense urban cluster), and
rural (very low density rural, low density rural and rural cluster)
\citep{florczyk2019ghsl}.

\subsection{Computation of population-level displacement
estimates}\label{sec-methods2.2}

\textbf{Estimating Internal Displacement.} We obtain population-level
estimates of internal displacement by correcting population counts
derived from the identified home location based on our smartphone GPS
data, to make them representative of the overall population. That is, we
correct mobile phone-derived population estimates to account for
differences in the use of mobile phone technology across locations in
Ukraine and over time. To this end, we adapted a deterministic model
proposed by Leasure and colleagues \citep{leasure2023nowcasting}.
Intuitively the approach involves first establishing our baseline
population; that is the pre-war population of Ukraine. We use population
data from WorldPop for 2020. Second, we identify the baseline number of
mobile phone users in Ukraine before the start of the full-scale
invasion by aggregating the number of unique devices in each home
location based on our GPS mobile phone data. Third, these two sets of
baseline estimates are used to compute the baseline mobile phone
penetration rate in each location \(i\) before the start of the
full-scale invasion (\(t=0\)). Formally, this rate can be expressed as:

\begin{equation}\phantomsection\label{eq-equation1}{ \psi_{i,t=0} = \frac {S_{i,t=0}} {N_{i,t=0}}}\end{equation}

where: \(S\) is the baseline median daily active mobile phone users
between January 1, 2022 to February 24, 2022; \(N\) is the baseline
total population in 2020 obtained from WorldPop.

Next, we estimate the present population \(N\) in location \(i\) at a
given point in time \(t\) from our GPS mobile phone data adjusting for
rate of mobile phone penetration. We do this by dividing the current
median daily active mobile phone users \(S\) at location \(i\) and time
\(t\) over the baseline smartphone penetration rate \(\psi\) at location
\(i\), assuming constant penetration rate since before the conflict:

\begin{equation}\phantomsection\label{eq-equation2}{ N_{i,t} = \frac {S_{i,t}} {\psi_{i}} }\end{equation}

We introduce a scaling factor (\(\theta\)) to account for potential
changes in mobile phone penetration rate over time as people leave
Ukraine and the local population shrinks. For this, we use daily
population counts of refugees from UNHCR and compute the net balance of
people leaving Ukraine which is denoted as \(R\). We compute the scaling
factor at time \(t\) using:

\begin{equation}\phantomsection\label{eq-equation3}{ \theta_{t} = \frac{\sum_{i=1}^I N_{i,t=0} - R_{t}}{\sum_{i=1}^I N_{i,t}} }\end{equation}

We use this scaling factor to compute our adjusted present population
(\(\hat{N}\)) multiplying the present population \(N\) obtained from
Equation~\ref{eq-equation2} and \(\theta\):

\begin{equation}\phantomsection\label{eq-equation4}{ \hat{N}_{i,t} = \theta_{t} N_{i,t} }\end{equation}

Using these population estimates, we can estimate changes in local
population across Ukraine over time. We can do this by subtracting our
adjusted current population estimates from the baseline population:

\begin{equation}\phantomsection\label{eq-equation5}{ \Delta_{i,t} = \hat{N}_{i,t} - N_{i,t=0}}\end{equation}

where: \(\Delta_{i,t}\) is the difference in population in location
\(i\) at time \(t\). For this difference, negative scores indicate a
local population loss due to internal population displacement, relative
to the baseline pre-war population. Similarly, positive scores indicate
a local population gain due to internal population displacement.

\textbf{Validation.} We assess the accuracy of our estimates of
population displacement. We measure the strength of the correspondence
between our set of mobile phone-based estimates \emph{versus} the
survey-based estimates produced by the IOM \citep{IOM2022}, and the
Facebook-based estimates produced by Leasure and colleagues
\citep{leasure2023nowcasting}. We do not anticipate a perfect linear
relationship as differences exist in the data source and methodologies
to produce the estimates. However, we do expect a high degree of
correlation indicating a high degree of temporal correspondence between
estimates. We compared against published IOM estimates at the national
and regional levels during the most of our analysis, and also
oblast-level estimates published by Leasure and colleagues.
Unfortunately more comparable spatially granular estimates at the raion
level were not available for our period of analysis. Supplementary Table
1 and Figures 2-4 display a set of Pearson correlation coefficients,
scatter plots and temporal relationships between these and our
estimates. The results reveal a high degree of geographic and temporal
correspondence between our estimates and those produced by IOM and
Leasure and colleagues across the set of metrics.

\subsection{Displacement metrics}\label{sec-metrics}

\textbf{Net balance of displacement.} To identify areas of high internal
population displacement, we analyse temporal changes in the net balance
of internal displacement. We compute the weekly average net balance of
internal displacements (\(NET\)) as the subtraction of the number of
people arriving (\(IN\)) minus the number of people leaving an area
(\(OUT\)). Positive scores indicate a net balance population gain due to
internal displacement, while negative scores denote a net balance
population loss. The net balance of internal displacement is computed
as:

\begin{equation}\phantomsection\label{eq-equation6}{ NET_{i,t} = {IN_{i,t}} - {OUT_{i,t}} }\end{equation}

\textbf{Population change across the urban-rural hierarchy.} We analyse
changes in population across the urban-rural hierarchy. We hypothesise
that dense urban areas have attracted a disproportionate number of
internally displaced people as they tend to serve as centres of
protective infrastructure and services for civilians. At the same time,
we expect decreases in population due to internal displacement in rural
areas. To determine the extent of population changes, we compute the
percentage change in population in individual areas, relative to the
baseline population (Equation~\ref{eq-equation7}). For this, we
spatially aggregated our internal displacement estimates at raion to
settlement areas based on the GHSL degree of urbanisation classification
described in Section (Section~\ref{sec-data-sources}).

\begin{equation}\phantomsection\label{eq-equation7}{ percent_{i,t} = \frac {\hat{N}_{i,t}} {N_{i,t0}} * 100 }\end{equation}

\textbf{Estimating returns.} We estimate the number of people who
returned to their usual place of residence before the start of the
full-scale invasion after a move somewhere else in the country. We
produce estimates at the raion and oblast level. We implemented and
compared two different approaches to estimate the number of returnees.
These approaches are based on the methods used by: (1) the IOM
\citep{iom2024-return}; and, (2) Leasure and team
\citep{leasure2023nowcasting}.

Similar to the former, we defined returns as those individual devices
which are recorded away for a period of at least two weeks and
subsequently in the same area identified as home location before the
start of the full-scale invasion. For individual areas, we computed the
relative proportion of return movements dividing the number of returns
relative to the total number of devices in a given home location. These
proportions are then multiplied by the pre-war baseline population to
produce population-level estimates of returnees.

Similar to Leasure and team, we defined returns as those individual
devices which are recorded away for an average of nine weeks and
subsequently in the same area identified as home location before the
start of the full-scale invasion (see Figure 6 in SM). Following the
same rationale explained for deriving our population-level estimates of
population displacement above, we divided the count of returns by the
pre-war smartphone penetration rate for each area. Acknowledging that
pre- and post-war smartphone penetration rates might differ as people
leaving Ukraine, we adjusted our estimates with scaling factors that
accounted for refugees who had left the country using
Equation~\ref{eq-equation3} and Equation~\ref{eq-equation4}.

We compared both sets of estimates and considered that the second
approach provides more reasonable estimates (see Figure 5 in SM). It
produces population-level return estimates displaying an increasing
trend, within the range of 1 million in February 2022 and 4 million July
2022, whereas the IOM-like approach generates estimates suggesting a
decline in the number of returns from over 2 million to around 500
thousand. We consider a decline in the number of the number of IDP to be
unrealistic. As Russian troops withdrew from northern Ukraine, we have
visual and anecdotal evidence to suggest that people have tended to
return to Kiev and northern areas in June and July 2022.

\textbf{Distance of displacement.} We estimated how far people moved
from their home location. Human mobility research indicates that most
people move locally to neighbouring areas. We sought to explore if a
similar process occurs for forced movements. To this end, we measured
the distance distribution for return moves, computing the distance
between the home location and destination before a return move is
detected at the raion level. The median distance travelled is often used
for comparative analysis \citep{stillwell2016}. Additionally, we measure
the distance for all moves computing between the home location, each
temporary stop and location before a return move is identified, to
assess difference between the overall distance distribution and that of
return moves. We identified that the set of distances are quite similar
suggesting that people tend to directly move to their ``final''
destination without long periods of overnight stay in intermediate stops
(see Figure 1 in SM). We measured the Haversine distance in kilometres
based on centriod coordinates indicating the angular distance between
two points on the surface of a sphere.

\section{Data availability}\label{data-availability}

The main dataset for our analysis comprises GPS location data from
mobile phone applications. These data are GDPR compliant and were
legally obtained from a data aggregator. We cannot identify the data
aggregator or share the data in compliance of the terms of data sharing
and use, and a sign of nondisclosure agreement. The other data used in
our analysis are openly available online for download: WorldPop data can
be obtained from \url{https://www.worldpop.org}; UNHCR data on refugees
can be retrieved from
\url{https://data.unhcr.org/en/situations/ukraine}; administrative
boundaries for Ukraine can be downloaded from
\url{https://data.humdata.org} and \url{https://gadm.org}; and, GHSL can
be accessed via \url{https://human-settlement.emergency.copernicus.eu}.

\section{Code availability}\label{code-availability}

The code, and relevant description to replicate the analysis and results
reported in this article can be found in an open-access Github
repository registered on the Open Science Framework with DOI {[}To be
added for publication{]}. We adopted an open and reproducible research
approach based on the use of open software. We used the R language in
RStudio and followed best practices in geographic data science
\citep{arribas-bel2021}.

\renewcommand\refname{References}
  \bibliography{refs.bib}

\begin{thebibliography}{10}
\expandafter\ifx\csname url\endcsname\relax
  \def\url#1{\burl{#1}}\fi
\expandafter\ifx\csname urlprefix\endcsname\relax\def\urlprefix{URL }\fi
\providecommand{\bibinfo}[2]{#2}
\providecommand{\eprint}[2][]{\url{#2}}
\providecommand{\doi}[1]{\url{https://doi.org/#1}}
\bibcommenthead

\bibitem{blattman2010}
\bibinfo{author}{Blattman, C.} \& \bibinfo{author}{Miguel, E.}
\newblock \bibinfo{title}{Civil war}.
\newblock \emph{\bibinfo{journal}{Journal of Economic Literature}}
  \textbf{\bibinfo{volume}{48}}, \bibinfo{pages}{3--57} (\bibinfo{year}{2010}).
\newblock \urlprefix\url{http://dx.doi.org/10.1257/jel.48.1.3}.

\bibitem{unhcr2023}
\bibinfo{author}{{UNHCR}}.
\newblock \emph{\bibinfo{title}{Global trends. Forced displacement in 2022.}}
  (\bibinfo{publisher}{{United Nations High Commissioner for Refugees. UNHCR}},
  \bibinfo{address}{Copenhagen, Denmark}, \bibinfo{year}{2022}).
\newblock \urlprefix\url{https://www.unhcr.org/refugee-statistics}.

\bibitem{idmc2024}
\bibinfo{author}{{IDMC}}.
\newblock \emph{\bibinfo{title}{2024 Global Report on Internal Displacement
  {(GRID)}}}  (\bibinfo{publisher}{{Internal Displacement Monitoring Centre,
  IDMC}}, \bibinfo{address}{Geneva, Switzerland}, \bibinfo{year}{2024}).
\newblock \urlprefix\url{https://www.unhcr.org/refugee-statistics}.

\bibitem{UNHCR_mid2023}
\bibinfo{author}{UNHCR}.
\newblock \bibinfo{title}{Mid-year trends 2023} (\bibinfo{year}{2023}).
\newblock \urlprefix\url{https://www.unhcr.org/mid-year-trends-report-2023}.

\bibitem{rowe2022}
\bibinfo{author}{Rowe, F.}
\newblock \bibinfo{title}{Using digital footprint data to monitor human
  mobility and support rapid humanitarian responses}.
\newblock \emph{\bibinfo{journal}{Regional Studies, Regional Science}}
  \textbf{\bibinfo{volume}{9}}, \bibinfo{pages}{665--668}
  (\bibinfo{year}{2022}).
\newblock \urlprefix\url{http://dx.doi.org/10.1080/21681376.2022.2135458}.

\bibitem{gonzález-leonardo2024}
\bibinfo{author}{{González{-}Leonardo}, M.}, \bibinfo{author}{Neville, R.},
  \bibinfo{author}{{Gil{-}Clavel}, S.} \& \bibinfo{author}{Rowe, F.}
\newblock \bibinfo{title}{Where have ukrainian refugees gone? identifying
  potential settlement areas across european regions integrating digital and
  traditional geographic data}.
\newblock \emph{\bibinfo{journal}{Population, Space and Place}}
  (\bibinfo{year}{2024}).
\newblock \urlprefix\url{http://dx.doi.org/10.1002/psp.2790}.

\bibitem{sarzin2017stocktaking}
\bibinfo{author}{Sarzin, Z.~I.}
\newblock \bibinfo{title}{Stocktaking of global forced displacement data}.
\newblock \emph{\bibinfo{journal}{World Bank Policy Research Working Paper}}
  (\bibinfo{year}{2017}).

\bibitem{tai2022}
\bibinfo{author}{Tai, X.~H.}, \bibinfo{author}{Mehra, S.} \&
  \bibinfo{author}{Blumenstock, J.~E.}
\newblock \bibinfo{title}{Mobile phone data reveal the effects of violence on
  internal displacement in afghanistan}.
\newblock \emph{\bibinfo{journal}{Nature Human Behaviour}}
  \textbf{\bibinfo{volume}{6}}, \bibinfo{pages}{624--634}
  (\bibinfo{year}{2022}).
\newblock \urlprefix\url{http://dx.doi.org/10.1038/s41562-022-01336-4}.

\bibitem{rowe2023}
\bibinfo{author}{Rowe, F.}
\newblock \emph{\bibinfo{title}{Big data}}, \bibinfo{pages}{42--47}
  (\bibinfo{publisher}{Edward Elgar Publishing}, \bibinfo{year}{2023}).
\newblock \urlprefix\url{http://dx.doi.org/10.4337/9781800883499.ch09}.

\bibitem{guideto2019}
\bibinfo{editor}{Salah, A.~A.}, \bibinfo{editor}{Pentland, A.},
  \bibinfo{editor}{Lepri, B.} \& \bibinfo{editor}{{Letouzé}, E.} (eds)
  \emph{\bibinfo{title}{Guide to Mobile Data Analytics in Refugee Scenarios}}
  (\bibinfo{publisher}{Springer International Publishing},
  \bibinfo{year}{2019}).
\newblock \urlprefix\url{http://dx.doi.org/10.1007/978-3-030-12554-7}.

\bibitem{drouhot2022}
\bibinfo{author}{Drouhot, L.~G.}, \bibinfo{author}{Deutschmann, E.},
  \bibinfo{author}{Zuccotti, C.~V.} \& \bibinfo{author}{Zagheni, E.}
\newblock \bibinfo{title}{Computational approaches to migration and integration
  research: promises and challenges}.
\newblock \emph{\bibinfo{journal}{Journal of Ethnic and Migration Studies}}
  \textbf{\bibinfo{volume}{49}}, \bibinfo{pages}{389--407}
  (\bibinfo{year}{2022}).
\newblock \urlprefix\url{http://dx.doi.org/10.1080/1369183X.2022.2100542}.

\bibitem{understa2011}
\bibinfo{author}{Hoglund, K.} \& \bibinfo{author}{Oberg, M.}
\newblock \emph{\bibinfo{title}{Understanding Peace Research Methods and
  Challenges}}  (\bibinfo{publisher}{Routledge}, \bibinfo{year}{2011}).
\newblock \urlprefix\url{http://dx.doi.org/10.4324/9780203828557}.

\bibitem{salehyan2015}
\bibinfo{author}{Salehyan, I.}
\newblock \bibinfo{title}{Best practices in the collection of conflict data}.
\newblock \emph{\bibinfo{journal}{Journal of Peace Research}}
  \textbf{\bibinfo{volume}{52}}, \bibinfo{pages}{105--109}
  (\bibinfo{year}{2015}).
\newblock \urlprefix\url{http://dx.doi.org/10.1177/0022343314551563}.

\bibitem{checchi2013validity}
\bibinfo{author}{Checchi, F.}, \bibinfo{author}{Stewart, B.~T.},
  \bibinfo{author}{Palmer, J.~J.} \& \bibinfo{author}{Grundy, C.}
\newblock \bibinfo{title}{Validity and feasibility of a satellite imagery-based
  method for rapid estimation of displaced populations}.
\newblock \emph{\bibinfo{journal}{International journal of health geographics}}
  \textbf{\bibinfo{volume}{12}}, \bibinfo{pages}{1--12} (\bibinfo{year}{2013}).
\newblock \urlprefix\url{https://doi.org/10.1186/1476-072X-12-4}.

\bibitem{nyhan2016exposure}
\bibinfo{author}{Nyhan, M.} \emph{et~al.}
\newblock \bibinfo{title}{{“Exposure Track” - The} impact of
  mobile-device-based mobility patterns on quantifying population exposure to
  air pollution}.
\newblock \emph{\bibinfo{journal}{Environmental science \& technology}}
  \textbf{\bibinfo{volume}{50}}, \bibinfo{pages}{9671--9681}
  (\bibinfo{year}{2016}).
\newblock \urlprefix\url{https://doi.org/10.1021/acs.est.6b02385}.

\bibitem{dewulf2016dynamic}
\bibinfo{author}{Dewulf, B.} \emph{et~al.}
\newblock \bibinfo{title}{Dynamic assessment of exposure to air pollution using
  mobile phone data}.
\newblock \emph{\bibinfo{journal}{International journal of health geographics}}
  \textbf{\bibinfo{volume}{15}}, \bibinfo{pages}{1--14} (\bibinfo{year}{2016}).
\newblock \urlprefix\url{https://doi.org/10.1186/s12942-016-0042-z}.

\bibitem{huang2019transport}
\bibinfo{author}{Huang, H.}, \bibinfo{author}{Cheng, Y.} \&
  \bibinfo{author}{Weibel, R.}
\newblock \bibinfo{title}{Transport mode detection based on mobile phone
  network data: {A} systematic review}.
\newblock \emph{\bibinfo{journal}{Transportation Research Part C: Emerging
  Technologies}} \textbf{\bibinfo{volume}{101}}, \bibinfo{pages}{297--312}
  (\bibinfo{year}{2019}).
\newblock \urlprefix\url{https://doi.org/10.1016/j.trc.2019.02.008}.

\bibitem{kim2023mobile}
\bibinfo{author}{Kim, J.~Y.}, \bibinfo{author}{Kubo, T.} \&
  \bibinfo{author}{Nishihiro, J.}
\newblock \bibinfo{title}{Mobile phone data reveals spatiotemporal recreational
  patterns in conservation areas during the {COVID} pandemic}.
\newblock \emph{\bibinfo{journal}{Scientific Reports}}
  \textbf{\bibinfo{volume}{13}}, \bibinfo{pages}{20282} (\bibinfo{year}{2023}).
\newblock \urlprefix\url{https://doi.org/10.1038/s41598-023-47326-y}.

\bibitem{lu2016unveiling}
\bibinfo{author}{Lu, X.} \emph{et~al.}
\newblock \bibinfo{title}{Unveiling hidden migration and mobility patterns in
  climate stressed regions: {A} longitudinal study of six million anonymous
  mobile phone users in {Bangladesh}}.
\newblock \emph{\bibinfo{journal}{Global Environmental Change}}
  \textbf{\bibinfo{volume}{38}}, \bibinfo{pages}{1--7} (\bibinfo{year}{2016}).
\newblock \urlprefix\url{https://doi.org/10.1016/j.gloenvcha.2016.02.002}.

\bibitem{grantz2020use}
\bibinfo{author}{Grantz, K.~H.} \emph{et~al.}
\newblock \bibinfo{title}{The use of mobile phone data to inform analysis of
  {COVID-19} pandemic epidemiology}.
\newblock \emph{\bibinfo{journal}{Nature communications}}
  \textbf{\bibinfo{volume}{11}}, \bibinfo{pages}{4961} (\bibinfo{year}{2020}).
\newblock \urlprefix\url{https://doi.org/10.1038/s41467-020-18190-5}.

\bibitem{shibuya2024}
\bibinfo{author}{Shibuya, Y.}, \bibinfo{author}{Jones, N.} \&
  \bibinfo{author}{Sekimoto, Y.}
\newblock \bibinfo{title}{Assessing internal displacement patterns in ukraine
  during the beginning of the russian invasion in 2022}.
\newblock \emph{\bibinfo{journal}{Scientific Reports}}
  \textbf{\bibinfo{volume}{14}} (\bibinfo{year}{2024}).
\newblock \urlprefix\url{http://dx.doi.org/10.1038/s41598-024-59814-w}.

\bibitem{leasure2023nowcasting}
\bibinfo{author}{Leasure, D.~R.} \emph{et~al.}
\newblock \bibinfo{title}{Nowcasting daily population displacement in {Ukraine}
  through social media advertising data}.
\newblock \emph{\bibinfo{journal}{Population and Development Review}}
  \textbf{\bibinfo{volume}{49}}, \bibinfo{pages}{231--254}
  (\bibinfo{year}{2023}).
\newblock \urlprefix\url{https://doi.org/10.1111/padr.12558}.

\bibitem{ranjan2012}
\bibinfo{author}{Ranjan, G.}, \bibinfo{author}{Zang, H.},
  \bibinfo{author}{Zhang, Z.-L.} \& \bibinfo{author}{Bolot, J.}
\newblock \bibinfo{title}{Are call detail records biased for sampling human
  mobility?}
\newblock \emph{\bibinfo{journal}{ACM SIGMOBILE Mobile Computing and
  Communications Review}} \textbf{\bibinfo{volume}{16}},
  \bibinfo{pages}{33--44} (\bibinfo{year}{2012}).
\newblock \urlprefix\url{http://dx.doi.org/10.1145/2412096.2412101}.

\bibitem{zhao2016}
\bibinfo{author}{Zhao, Z.} \emph{et~al.}
\newblock \bibinfo{title}{Understanding the bias of call detail records in
  human mobility research}.
\newblock \emph{\bibinfo{journal}{International Journal of Geographical
  Information Science}} \textbf{\bibinfo{volume}{30}},
  \bibinfo{pages}{1738--1762} (\bibinfo{year}{2016}).
\newblock \urlprefix\url{http://dx.doi.org/10.1080/13658816.2015.1137298}.

\bibitem{pestre2019}
\bibinfo{author}{Pestre, G.}, \bibinfo{author}{{Letouzé}, E.} \&
  \bibinfo{author}{Zagheni, E.}
\newblock \bibinfo{title}{The abcde of big data: Assessing biases in
  call-detail records for development estimates}.
\newblock \emph{\bibinfo{journal}{The World Bank Economic Review}}
  \textbf{\bibinfo{volume}{34}}, \bibinfo{pages}{S89--S97}
  (\bibinfo{year}{2019}).
\newblock \urlprefix\url{http://dx.doi.org/10.1093/wber/lhz039}.

\bibitem{grantz2020}
\bibinfo{author}{Grantz, K.~H.} \emph{et~al.}
\newblock \bibinfo{title}{The use of mobile phone data to inform analysis of
  covid-19 pandemic epidemiology}.
\newblock \emph{\bibinfo{journal}{Nature Communications}}
  \textbf{\bibinfo{volume}{11}} (\bibinfo{year}{2020}).
\newblock \urlprefix\url{http://dx.doi.org/10.1038/s41467-020-18190-5}.

\bibitem{IOM2022}
\bibinfo{author}{IOM}.
\newblock \bibinfo{title}{Ukraine internal displacement report - general
  population survey - round 11 (25 november - 5 december 2022)}.
\newblock \bibinfo{type}{Tech. Rep.}, \bibinfo{institution}{International
  Organization for Migration (IOM)} (\bibinfo{year}{2022}).

\bibitem{walker2024}
\bibinfo{author}{Walker, N.}
\newblock \bibinfo{title}{Conflict in {Ukraine: A} timeline (current conflict,
  2022-present)}.
\newblock \emph{\bibinfo{journal}{{House of Commons Library}}}
  \urlprefix\url{https://commonslibrary.parliament.uk/research-briefings/cbp-9847/}.

\bibitem{yana2021}
\bibinfo{author}{Dlugy, Y.}
\newblock \bibinfo{title}{The fall of {Sievierodonetsk}}.
\newblock \emph{\bibinfo{journal}{{The New York Times}}}
  \urlprefix\url{https://www.nytimes.com/2022/06/24/briefing/russia-ukraine-war-sievierodonetsk-romania.html}.

\bibitem{rowe2022b}
\bibinfo{author}{Rowe, F.}, \bibinfo{author}{Neville, R.} \&
  \bibinfo{author}{{González-Leonardo}, M.}
\newblock \bibinfo{title}{Sensing population displacement from ukraine using
  facebook data: Potential impacts and settlement areas}.
\newblock \emph{\bibinfo{journal}{OSF preprint}}  (\bibinfo{year}{2022}).
\newblock \urlprefix\url{http://dx.doi.org/10.31219/osf.io/7n6wm}.

\bibitem{iom2023-return}
\bibinfo{author}{Galindo, J.}
\newblock \bibinfo{title}{{Return, reintegration and recovery. IOM’s position
  on returns to Ukraine}}.
\newblock \bibinfo{type}{Tech. Rep.}, \bibinfo{institution}{{International
  Organization for Migration (IOM)}} (\bibinfo{year}{2023}).

\bibitem{iom2024-return}
\bibinfo{author}{IOM}.
\newblock \bibinfo{title}{{Ukraine returns report. General population survey.
  Round 16}}.
\newblock \bibinfo{type}{Tech. Rep.}, \bibinfo{institution}{{International
  Organization for Migration (IOM)}} (\bibinfo{year}{2024}).

\bibitem{stillwell2016}
\bibinfo{author}{Stillwell, J.} \emph{et~al.}
\newblock \bibinfo{title}{Internal migration around the world: comparing
  distance travelled and its frictional effect}.
\newblock \emph{\bibinfo{journal}{Environment and Planning A: Economy and
  Space}} \textbf{\bibinfo{volume}{48}}, \bibinfo{pages}{1657--1675}
  (\bibinfo{year}{2016}).
\newblock \urlprefix\url{http://dx.doi.org/10.1177/0308518X16643963}.

\bibitem{IOM2022r1}
\bibinfo{author}{IOM}.
\newblock \bibinfo{title}{Ukraine — idp figures: General population survey,
  round 1 (9 - 16 march 2022)}.
\newblock \bibinfo{type}{Tech. Rep.}, \bibinfo{institution}{International
  Organization for Migration (IOM)} (\bibinfo{year}{2022}).

\bibitem{graells-garrido2021}
\bibinfo{author}{Graells-Garrido, E.}, \bibinfo{author}{Serra-Burriel, F.},
  \bibinfo{author}{Rowe, F.}, \bibinfo{author}{Cucchietti, F.~M.} \&
  \bibinfo{author}{Reyes, P.}
\newblock \bibinfo{title}{A city of cities: Measuring how 15-minutes urban
  accessibility shapes human mobility in barcelona}.
\newblock \emph{\bibinfo{journal}{PLOS ONE}} \textbf{\bibinfo{volume}{16}},
  \bibinfo{pages}{e0250080} (\bibinfo{year}{2021}).
\newblock \urlprefix\url{http://dx.doi.org/10.1371/journal.pone.0250080}.

\bibitem{Araietal2022}
\bibinfo{author}{Magpantay, E.} \emph{et~al.}
\newblock \bibinfo{title}{Methodological guide on the use of mobile phone data:
  Measuring the information society}.
\newblock \emph{\bibinfo{journal}{New York: United Nations Statistics
  Division}}  (\bibinfo{year}{2022}).

\bibitem{worldpopukraine}
\bibinfo{author}{{WorldPop}}.
\newblock \bibinfo{title}{Ukraine conflict - {WorldPop}}
  \bibinfo{note}{Available at: \url{https://www.worldpop.org/events/ukraine/}.
  [Accessed 3 Sep. 2023].}

\bibitem{OPD}
\bibinfo{author}{{UNHCR}}.
\newblock \bibinfo{title}{Ukraine refugee situation}  (\bibinfo{year}{2024}).
\newblock \bibinfo{note}{Available at:
  \url{https://data.unhcr.org/en/situations/ukraine}. [Accessed 3 Oct. 2024].}

\bibitem{GADM}
\bibinfo{author}{GADM}.
\newblock \bibinfo{title}{{GADM} data: the database of global administrative
  areas} (\bibinfo{year}{2022}).
\newblock \bibinfo{note}{Available at: \url{https://gadm.org/data.html}.
  [Accessed 15 Jan. 2024].}

\bibitem{florczyk2019ghsl}
\bibinfo{author}{Florczyk, A.~J.} \emph{et~al.}
\newblock \bibinfo{title}{{GHSL data package 2019. Public release GHS P2019}}.
\newblock \emph{\bibinfo{journal}{{Luxembourg: Publications Office of the
  European Union}}} \textbf{\bibinfo{volume}{29788}}, \bibinfo{pages}{290498}
  (\bibinfo{year}{2019}).

\bibitem{arribas-bel2021}
\bibinfo{author}{Arribas-Bel, D.}, \bibinfo{author}{Green, M.},
  \bibinfo{author}{Rowe, F.} \& \bibinfo{author}{Singleton, A.}
\newblock \bibinfo{title}{Open data products-a framework for creating valuable
  analysis ready data}.
\newblock \emph{\bibinfo{journal}{Journal of Geographical Systems}}
  \textbf{\bibinfo{volume}{23}}, \bibinfo{pages}{497--514}
  (\bibinfo{year}{2021}).
\newblock \urlprefix\url{http://dx.doi.org/10.1007/s10109-021-00363-5}.

\end{thebibliography}

\end{document}